# Dynamical Structure Factor of Fulde-Ferrell-Larkin-Ovchinnikov Superconductors

Arghya Dutta[1] and Jayanta K. Bhattacharjee[1,2]

*(1) S. N. Bose National Centre for Basic Sciences, JD-Block, Sector-III, Salt Lake, Kolkata-700098, India,
email:arghya@bose.res.in*
*(2) Harish-Chandra Research Institute, Chhatnag Road, Jhusi, Allahabad-211019, India*

**Abstract.** Superconductor with a spatially-modulated order parameter is known as Fulde-Ferrell-Larkin-Ovchinnikov (FFLO) superconductor. Using the time-dependent Ginzburg-Landau (TDGL) formalism we have theoretically studied the temporal behaviour of the equal-time correlation function, or the structure factor, of a FFLO superconductor following a sudden quench from the unpaired, or normal, state to the FFLO state. We find that quenching into the ordered FFLO phase can reveal the existence of a line in the mean-field phase diagram which cannot be accessed by static properties.

**Keywords:** Inhomogeneous superconductors and superconducting systems, Superconductivity phase diagrams, Phenomenological theories (two-fluid, Ginzburg-Landau, etc.) .

**PACS:** 74.81.-g, 74.25.Dw, 74.20.De

## INTRODUCTION

The inhomogeneous superconductivity, proposed by Fulde and Ferrell and by Larkin and Ovchinnikov (FFLO) [1] almost 60 years ago, remains a very active field of theoretical and experimental research even today. The reason for this is, though predicted so many years ago, the conclusive experimental evidence of FFLO superconducting state is still lacking. The FFLO state is a superconducting state with finite-momenta Cooper pairs, as opposed to zero-momenta Cooper pairs of Bardeen-Cooper-Schrieffer (BCS) superconductors, and this results in a spatially-modulated superconducting order parameter, which is the main signature of FFLO state. The FFLO state occurs in imbalanced two-component fermionic systems, more specifically, FFLO state has been predicted to occur in heavy-fermion superconductor $CeCoIn_5$, in layered organic superconductors, in imbalanced ultra-cold Fermi gases and, surprisingly, in compact stars, as quantum chromodynamics colour superconductivity.

Continuing this search for FFLO state, there have been some recent studies[2] which looked for the role of FFLO state when one suddenly quenches an imbalanced fermionic system from the normal, i.e. unpaired, state to the superconducting state and they found that following a quench, the FFLO, or, equivalently, finite-momenta superfluid state, occupies a major portion of the phase diagram in contrast to the narrow region occupied by the FFLO state in equilibrium phase diagram, which makes it very hard to find the FFLO state experimentally.

Encouraged by these studies, in this paper, we have studied the dynamical behaviour of the FFLO order parameter following a sudden normal-to-FFLO quench, both in early and late time limits. Specifically, we have calculated the equal-time dynamical correlation factor, or the structure factor, and find that quenching the system provides a handle to probe the mean-field phase diagram of this system and this dynamics reveals some features which are inaccessible to static analysis.

## THE MODEL AND RESULTS

To start with, we take the phenomenological Ginzburg-Landau (GL) free energy of a FFLO superconductor, following the notation of Casalbuoni et al [3], derived from the microscopic Hamiltonian, which reads

$$\Omega = \sum_{\vec{k}}(\alpha + \frac{2\beta}{3}k^2 + \frac{8\gamma}{15}k^4)|\varphi_{\vec{k}}|^2$$
$$+ \frac{1}{2}\sum_{\vec{k}_i}(\beta + \frac{4\gamma}{9}(k_1^2 + k_2^2 + k_3^2 + k_4^2 + \vec{k}_1 \cdot \vec{k}_3 + \vec{k}_2 \cdot \vec{k}_4))\varphi_{\vec{k}_1}\varphi_{\vec{k}_2}^*\varphi_{\vec{k}_3}\varphi_{\vec{k}_4}^*$$
$$+ \frac{\gamma}{3}\sum_{\vec{k}_i}\varphi_{\vec{k}_1}\varphi_{\vec{k}_2}^*\varphi_{\vec{k}_3}\varphi_{\vec{k}_4}^*\varphi_{\vec{k}_5}\varphi_{\vec{k}_6}^*.$$
(1)

In the above α is scaled temperature, β and γ are constants. φ is the momentum-dependent order parameter and both quartic and sextic terms have momentum-conserving delta functions. The mean-field phase diagram of this GL free energy [4] has been shown and explained in Fig.(1).

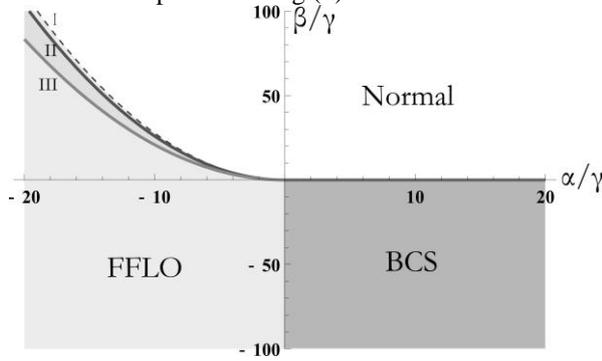

**FIGURE 1.** The mean-field phase diagram of a FFLO superconductor, plotted as a function of α/γ (or temperature) and β/γ. There are three phases: normal, or unpaired, phase, standard BCS superconducting phase and the inhomogeneous FFLO phase. When one lowers the temperature, in the negative-β region, the system encounters a metastable line (line I) and a first-order phase transition line (line II), respectively. On curve III, the coefficient of the quadratic term in order parameter vanishes.

Now to analyze the dynamical behavior of the system, we write down the TDGL equation which reads

$$\frac{\partial \varphi_{\vec{k}}}{\partial t} = -\Gamma \frac{\delta \Omega}{\delta \varphi_{-\vec{k}}} \quad \ldots (2)$$

The experimentally measurable quantity for this system is the structure factor which is defined as

$$S(\vec{k},t) = \langle \varphi_{\vec{k}}(t)\varphi_{-\vec{k}}(t) \rangle \quad \ldots (3)$$

With this definition, we have determined the structure factor in Hartree approximation with the help of Eq. (1) and (2). At early time limit, if $\alpha > 5\beta^2/24\gamma$ and we quench in the region between line II and III in the phase diagram all modes of the order parameter decays quickly. However if $\alpha < 5\beta^2/24\gamma$, there exist a band of wave numbers around a critical wave number ($k_C = \sqrt{(5|\beta|/8\gamma)}$) for which the structure factor grows. Thus the decay or growth of the structure factor in short time limit actually indicates the existence of the line III in the phase diagram, which static analysis cannot probe [4].

The critical wave number can be found experimentally if one quench the system below the first-order line and look for the long-time dynamics. In the long-time limit the structure factor can be calculated by performing a self-consistent analysis, in the sense that the system will order in the long-time limit, and the form the structure factor shows that all the modes of the order parameter decay expect the mode with the critical wave number. In short, if we quench an imbalanced fermionic system from the normal, or unpaired, state to the superconducting state in a region where β<0 (which can be set by tuning the temperature and the imbalance parameter, generally chemical potential difference between the fermionic species), then in the long-time limit the system will equilibrate to an inhomogeneous superconducting state with a particular wave number.

## CONCLUSIONS

In this paper we have analyzed the inhomogeneous superconducting ground state, the FFLO state, of an imbalanced fermionic system and find that while static properties, such as specific heat, cannot probe the complete phase diagram, the dynamical structure factor provides additional information about the phase diagram which has direct experimental significance in finding the elusive FFLO state. One possible extension of our analysis would be to perform the quench analysis beyond the Hartree approximation and study the non-linear effects of the higher order terms in the GL free energy.

## ACKNOWLEDGMENTS

One of the authors (A.D.) thanks Council of Scientific and Industrial Research, India for financial support in the form of fellowship (File No.09/575(0062)/2009-EMR-1). AD also thanks Harish-Chandra Research Institute for hospitality and support during visit.